\documentclass[sigconf]{acmart}

\AtBeginDocument{%
  \providecommand\BibTeX{{%
    \normalfont B\kern-0.5em{\scshape i\kern-0.25em b}\kern-0.8em\TeX}}}

\setcopyright{acmcopyright}
\copyrightyear{2022}
\acmYear{2022}
\acmDOI{https://doi.org/10.1145/3526113.3545685}

\acmConference[UIST '22]{ACM Symposium on User Interface Software and Technology}{October 29--November 2,
  2022}{Bend, OR}
\acmPrice{15.00}
\acmISBN{978-1-4503-9320-1/22/10}
\acmBooktitle{UIST '22: ACM Symposium on User Interface Software and Technology, October 29--November 2, 2022,Bend, OR}
\usepackage{color}
\newcommand{\commentout}[1]{}
\newcommand{\add}[1]{#1}
\newcommand{\whisp}[1]{$\mathbb{#1}$}

\usepackage{graphicx}

\begin{document}

\title{DualVoice: Speech Interaction that Discriminates between Normal and Whispered Voice Input}

\author{Jun Rekimoto}
\orcid{0000-0002-3629-2514}
\affiliation{%
  \institution{The University of Tokyo}
  \streetaddress{7-3-1, Hongo}
  \city{Bunkyo-ku}
  \state{Tokyo}
  \country{Japan}
  \postcode{113-0033}
}
\affiliation{%
  \institution{Sony CSL Kyoto}
  \streetaddress{13-1, Hontoro-cho, Shimogyo-ku}
  \city{Kyoto-shi}
  \state{Kyoto}
  \country{Japan}
  \postcode{600-8086}
}
\email{rekimoto@acm.org}

\renewcommand{\shortauthors}{Jun Rekimoto}

\begin{abstract}
Interactions based on automatic speech recognition (ASR) have become widely used, with speech input being increasingly utilized to create documents. However, as there is no easy way to distinguish between commands being issued and text required to be input in speech, misrecognitions are difficult to identify and correct, meaning that documents need to be manually edited and corrected. The input of symbols and commands is also challenging because these may be misrecognized as text letters. To address these problems, this study proposes a speech interaction method called DualVoice, by which commands can be input in a whispered voice and letters in a normal voice. The proposed method does not require any specialized hardware other than a regular microphone, enabling a complete hands-free interaction. The method can be used in a wide range of situations where speech recognition is already available, ranging from text input to mobile/wearable computing. Two neural networks were designed in this study, one for discriminating normal speech from whispered speech, and the second for recognizing whisper speech. A prototype of a text input system was then developed to show how normal and whispered voice can be used in speech text input. Other potential applications using DualVoice are also discussed.
\end{abstract}



\begin{teaserfigure}
\centering
\includegraphics[width=0.8\textwidth]{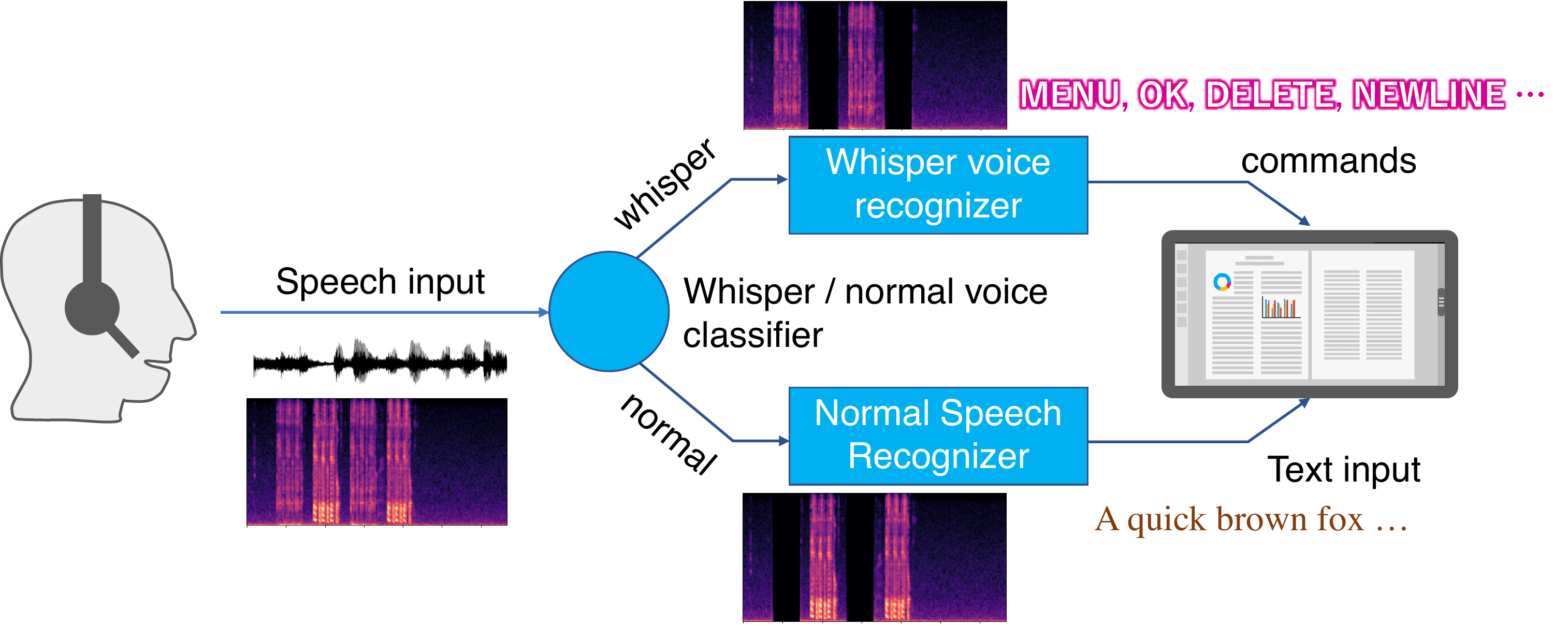}
\caption{DualVoice speech interaction: It uses ``whispering'' for commands whereas a normal voice is used for text inputs.}
\label{fig:teaser}
\Description[DualVoice Overview]{DualVoice Overview: it uses whispering for commands whereas a normal voice for text inputs.}
\end{teaserfigure}

\maketitle

\section{Introduction}

Automatic speech recognition (ASR) is used in a wide variety of  applications, ranging from techniques for text input to operating devices such as smart speakers, mobile devices, or car navigation systems. Voice input requires no specialized input device other than a microphone and thus can be operated hands-free. Furthermore, as text input based on speech can be much faster than typing, voice input can also be used to note ideas and quickly enter manuscript drafts.

However, automatic speech recognition is often inaccurate, with recognition errors that need to be corrected using traditional input methods, such as a keyboard and mouse, which eliminates the advantage of hands-free operation. In addition, misrecognition errors can arise when inputting commands or special characters other than regular text. For example, if you want to enter ``?'', you may say ``question mark,'' but it might be entered as the text itself, not the symbol, or, if you want to start a new paragraph and say ``new paragraph,'' it may be recognized literally and input as the text ``new paragraph''. Similar problems occur when commands such as ``delete word'' and ``new paragraph,'' which are frequently used in text editing, need to be followed.

Deciding in advance that saying ``new line'' always means the new-line command and not text input is a possible solution; however, when the string ``new line,'' needs to be entered, errors may occur. Moreover, users may find remembering all the commands and text input phrases difficult. The shift key on a regular keyboard may also be used to switch between modalities, but the hands-free advantage of voice input would be lost. 

To address these issues, this study proposes a speech interaction method called DualVoice, which uses ``whispering'' for commands and a normal voice for text inputs (Figure~\ref{fig:teaser}).

%\footnote{\add{A preliminary version of this work has been published as CHI EA Late-Breaking-Work (LBW)~\cite{10.1145/3491101.3519700}.}}

Whispering is a mode of speech that many users can implement without special training. Switching between whispering and a normal voice, users can give multiple meanings to speech. For example, saying ``new line'' in a normal voice can mean to input that string, whereas whispering ``new line'' can invokes a new-line command. Similarly, to correct a recognition error, users can whisper ``candidates,'' which would result in alternative recognition candidates being displayed; the user can then select one of them by whispering the number (``one,'' ``two,'' etc.) corresponding to that candidate.

The relationship between a normal voice and whisper can be compared to the relationship between text input and command input, where the conventional voice recognition method can be regarded as a keyboard that does not include command or symbol keys.
However, a keyboard used for computer interactions has function keys, command keys, keys for symbols, and other means of entering commands, as well as letter input. Thus, users can invoke those functions without explicitly changing the interaction mode or having to take their hands off the keyboard. Similarly, with the proposed DualVoice method, text and command input can co-exist without special mode conversion operations.

\add{
Interactions using whispering also have other possibilities not limited to commands when entering text. Voice interaction is difficult to use in public environments, where preservation of privacy and social acceptability become issues; a whisper voice input has the potential to solve these issues. If one can discriminate between normal voice (for normal conversation) and whisper voice for computer inputs, effective interaction for wearable, mobile, and virtual-environment computing may be possible.
}

\begin{figure}
\centering
\includegraphics[width=0.45\textwidth]{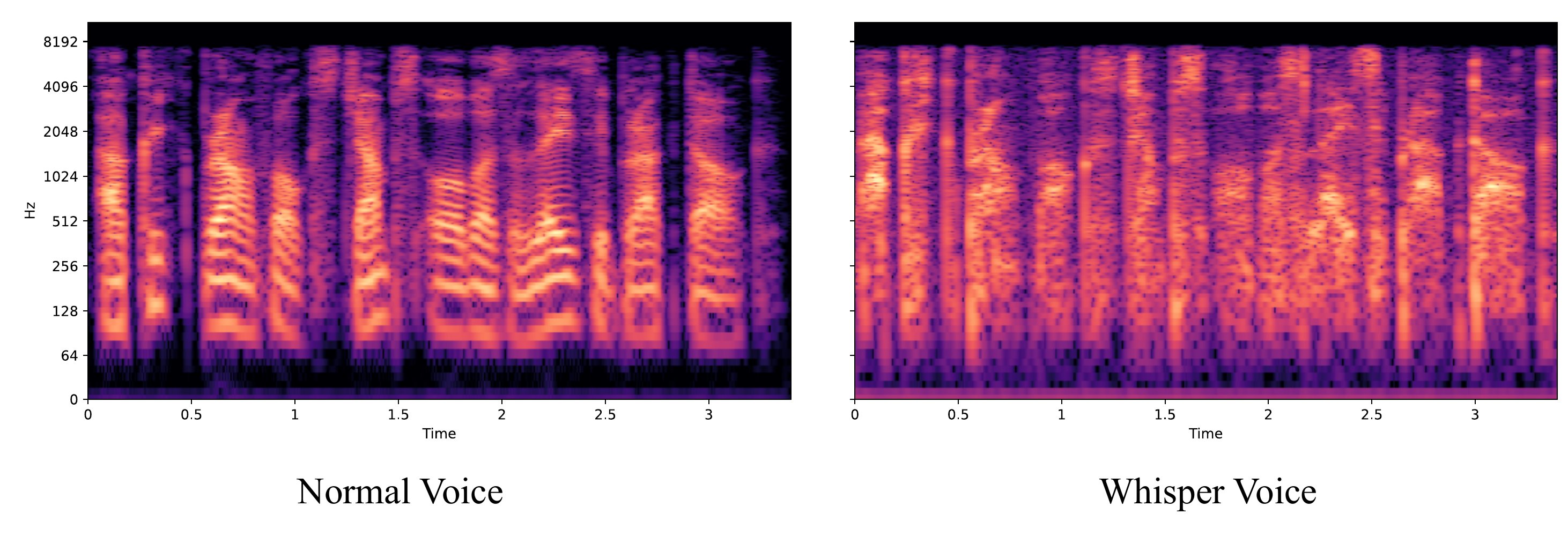}
\caption{Spectrograms of normal and whispered voices saying ``A quick brown fox jumps over the lazy black dog.'' }
\label{fig:whisper}
\Description[Spectrograms of normal and whisper voices]{Spectrograms of normal and whispered voices saying ``A quick brown fox jumps over the lazy black dog.''}
\end{figure}

Figure~\ref{fig:whisper} shows spectrograms of utterances with normal and whispered voices. To realize DualVoice interaction, two neural networks were developed for this study: one to distinguish between whispered and normal voices, and the other to recognize whispered utterances.

\add{
The contributions are summarized as follows:
\begin{itemize}
    \item Speech interfaces that discriminate between whisper and normal voices were constructed,
    \item A whisper--speech recognizer based on self-supervised learning was built, and
    \item A highly accurate neural network for whisper--normal speech discrimination was constructed.
\end{itemize}
}

\section{Related Work}

\begin{figure}
\centering
\includegraphics[width=0.43\textwidth]{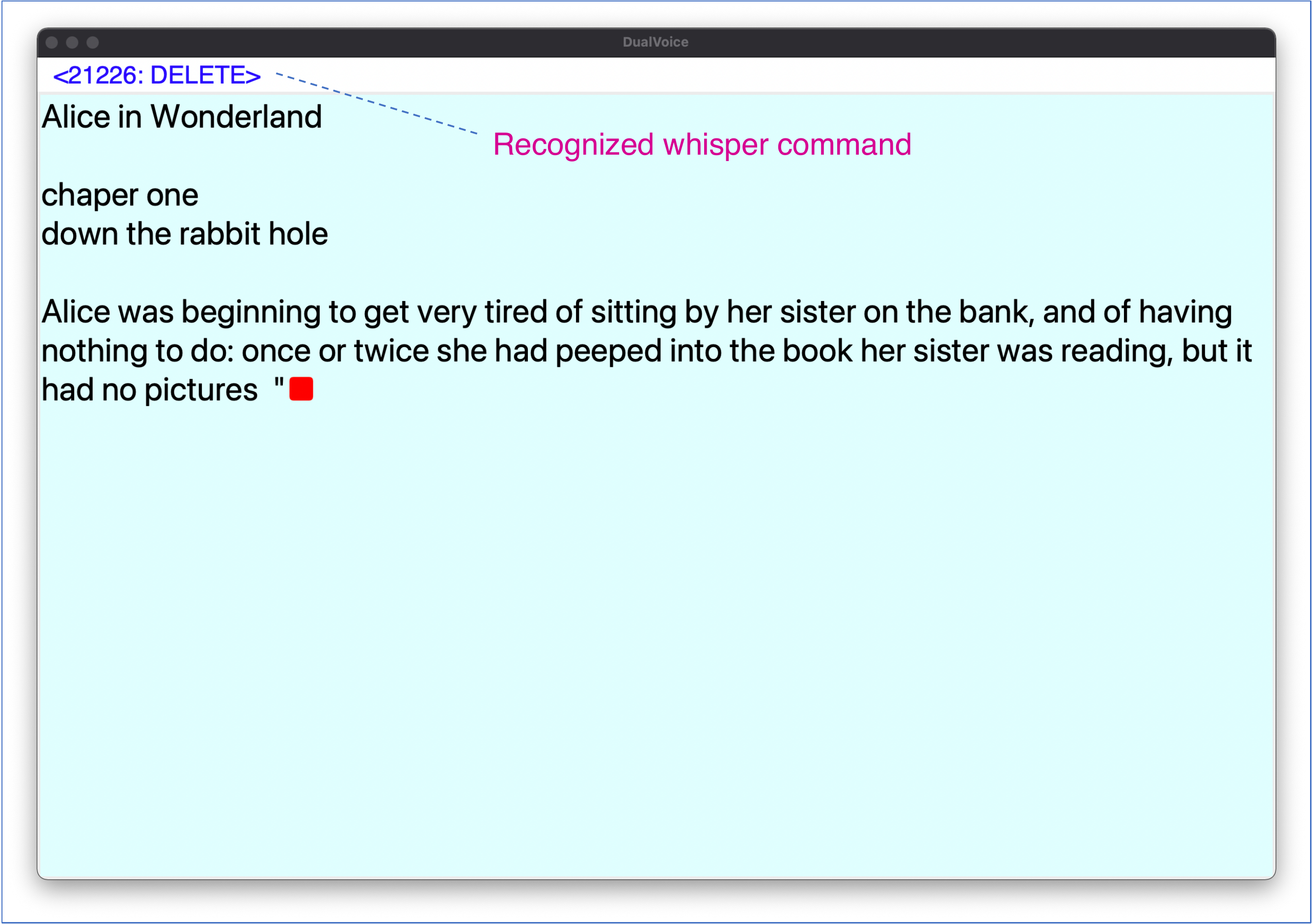}
\caption{DualVoice text input examples.}
\label{fig:textinput}
\Description[DualVoice text input]{DualVoice text input examples}
\end{figure}

\begin{figure*}
\centering
\includegraphics[width=0.95\textwidth]{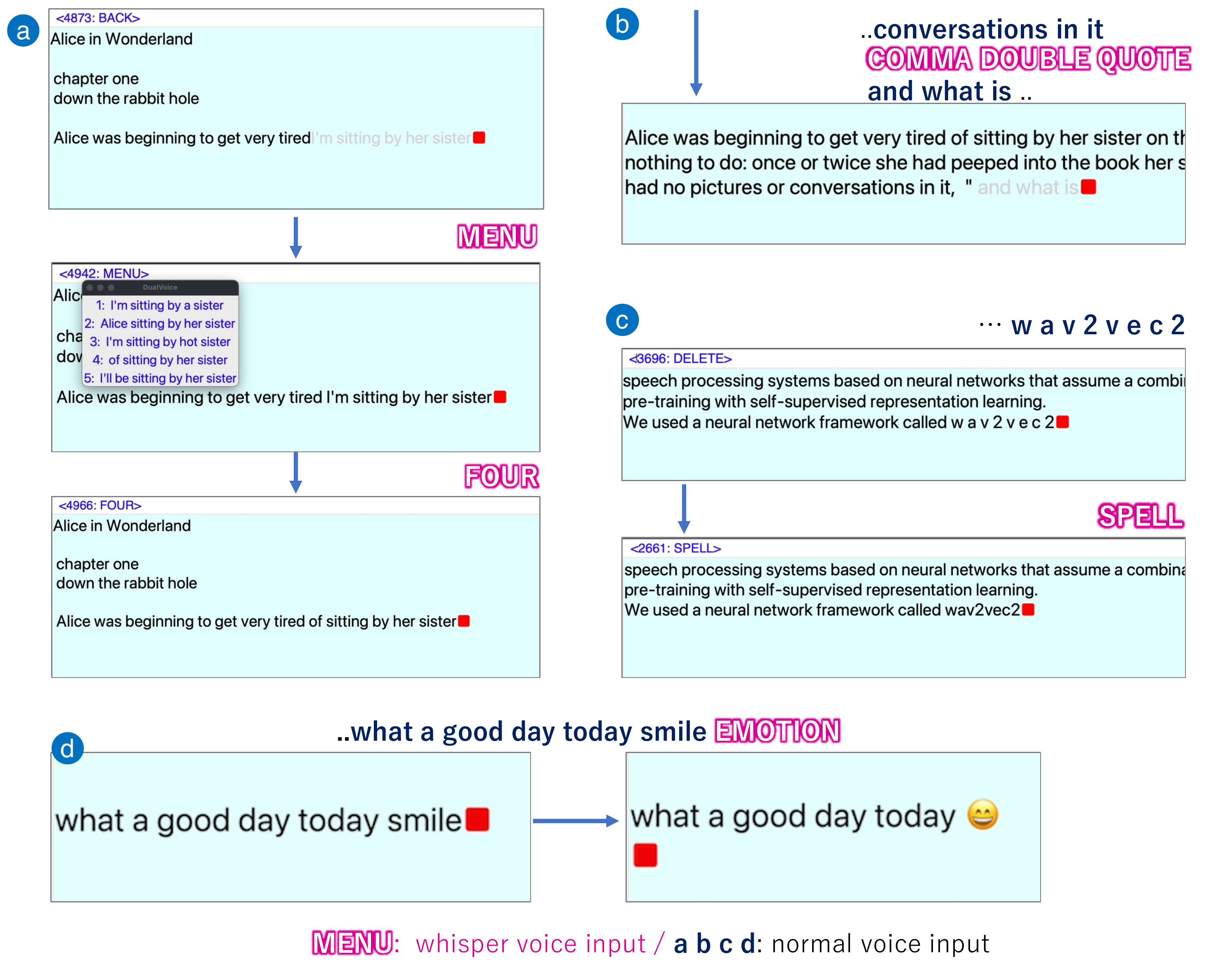}
\caption{Examples of DualVoice interactions: (a) Correction of recognition errors: whispering ``\whisp{MENU}'' invokes a menu for possible recognition candidates.  A user may select the required candidate by whispering the corresponding number. (b) Entering symbols by whispering ``\whisp{COMMA}'' and ``\whisp{DOUBLE}\ \whisp{QUOTE}.'' (c) Combining the spelled input by whispering ``\whisp{SPELL}.'' (d) Inputting Emoji by saying ``smile'' in a normal voice, followed by whispering ``\whisp{EMOTION}.''}
\label{fig:sample}
\Description[DualVoice interactions]{Examples of DualVoice interactions}
\end{figure*}

\begin{figure*}
\centering
\includegraphics[width=0.8\textwidth]{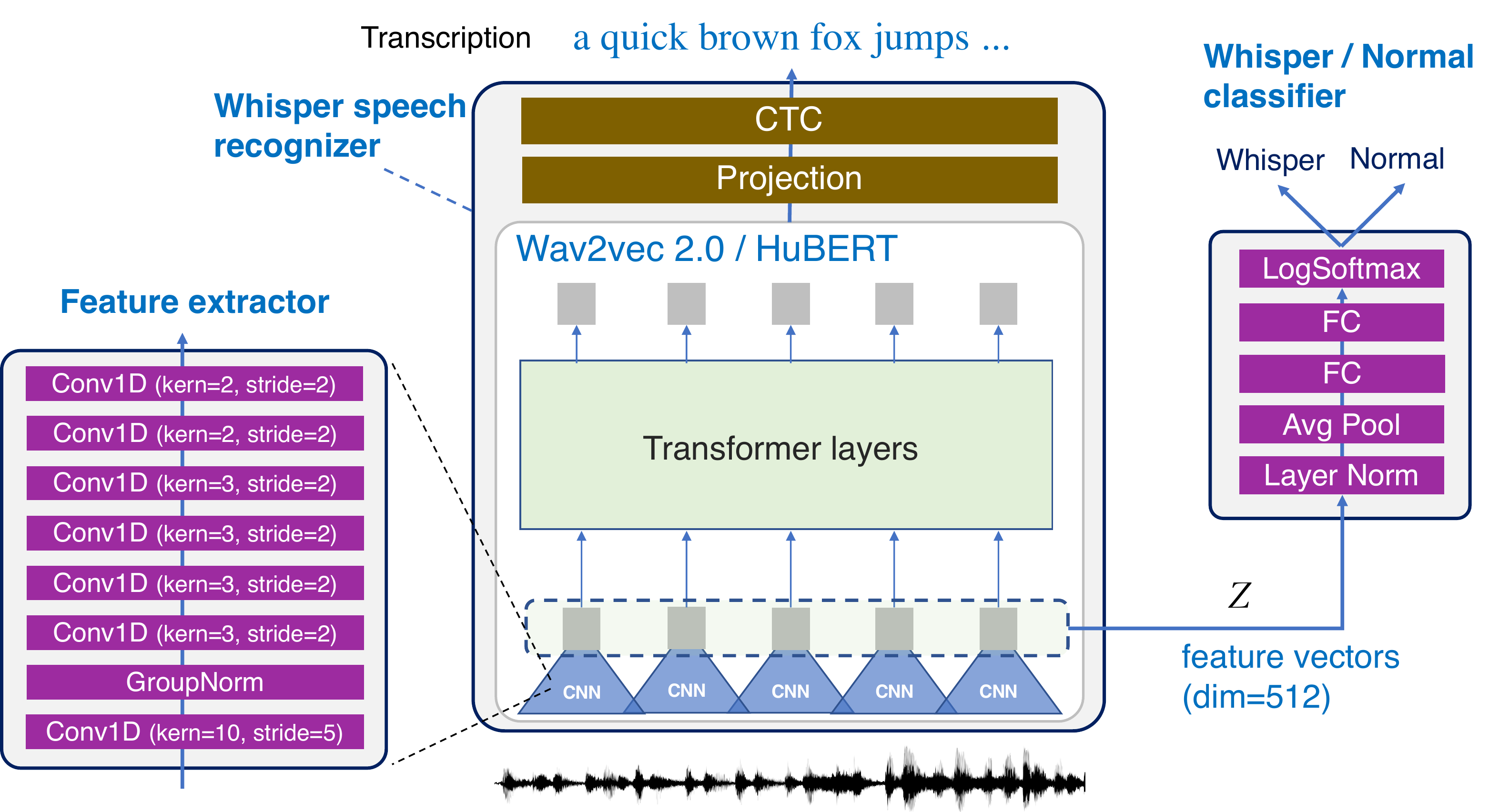}
\caption{Whisper voice recognition and whisper / normal classifier.}
\label{fig:wav2vec2asr}
\Description[Whisper/Normal recognition and classification]{Whisper voice recognition and whisper / normal classifier.}
\end{figure*}

Goto et al. proposed ``speech-shift,'' which specifies the mode of speech input by intentionally controlling the pitch of speech~\cite{Goto2003-lu}; i.e. if the fundamental frequency ($F_0$) of an utterance exceeds a specified threshold, it is judged as a different mode. This method introduced two modes of speech input; however, the user is required to speak at an unnaturally high pitch for stable recognition. In contrast, switching between whisper and normal speech is more natural, and can be done more clearly without setting the (user-unknown) threshold.

Goto et al. also developed a method that automatically detects filled (vocalized) pauses, which are hesitation phenomena of utterances in speech commands, and proposes possible candidates that can fill the command~\cite{Goto1999}. For instance, if the user says ``play, Beeee...'' and stops, the system detects ``eeee'' as the filled vocalized pauses of hesitation and proposes filling candidates such as ``Beatles'' or ``Beach Boys.'' This method demonstrates the possibility of indicating non-verbal intentions in speech; however, it can only utilize hesitation and not arbitrary commands as can be utilized by the proposed approach. Moreover, making vocalized pauses is only possible after vowels in speech, not after consonants.

PrivateTalk~\cite{privatetalk} uses a hand partially covering the mouth from one side for activating the speech command. Although the main objective of this research was to preserve privacy, this technique might also be used to distinguish normal speech (without hand covering) from command (with hand covering).  However, explicit hand gestures are required, and thus, the method can no longer be considered a ``hands free'' interaction. Furthermore, whereas PrivateTalk needs two microphones (attached to the left and right earphones) to recognize the effect of hand covering, the proposed approach only requires a single standard microphone.  To preserve privacy, this study also considered that it is more natural and effective to speak in whispers than to cover the mouth.

DualBreath~\cite{Onishi2021-tj} is a system that uses breathing as a command, distinguishing a command from normal exhalation by discriminating between inhaling and exhaling air through the nose and mouth simultaneously. This system can express triggers equivalent to a simple button press; however, it cannot express commands as rich as that expressed by whispers.
ProxiMic is a sensing technology that detects a user's utterance via a microphone device located close to the mouth~\cite{10.1145/3411764.3445687}. It is intended for use as a wake-up-free speech for smart speakers and similar voice-controlled devices. However, it requires physical motion (moving a microphone close to the mouth), and thus mixing normal and close-to-the-mouth utterances is not easy.

Fukumoto proposed SilentVoice, where a user can input speech using {\it ingressive speech}, an utterance made while inhaling (breathing-in)~\cite{10.1145/3242587.3242603}. Although the method is designed mainly for silent speech, it can also distinguish ingressive speech from normal speech. However, it requires a special microphone placed very close to the mouth, and training is needed for users to speak correctly in the ingressive mode. Moreover, frequently changing between normal and ingressive utterances is difficult.

\add{
There are a series of silent speech researches, where a user's silent utterances or silent commands are recognized with various sensing configurations including lip-reading, blowing, EMG, and ultrasound~\cite{10.1145/3172944.3172977,10.1145/3290605.3300376,10.1145/3242587.3242599,7310970,10.1145/2971763.2971765,10.1145/2634317.2634322}. Whisper speech has similar characteristics to silent speech, such as preserving social acceptability in public spaces; however, it has the potential to be more commonly used given that it can be recognized by ordinary microphones.
}

With some Speech Input Systems, such as Google Cloud Speech-to-Text~\cite{googlespeech}, one can obtain ``.'  input when you say ``period.'' However, to distinguish this utterance from the ``period' in the sentence, it is necessary to leave a certain amount of pause before and after it, which reduces the overall input speed. \add{It is conceivable to discriminate between text input and text normalization commands with a higher level of contextual understanding, such as using n-grams, but we consider our proposed method provides a more straightforward and more reliable means of discrimination.}

Research on whisper voice recognition~\cite{Denby:2010:SSI:1746726.1746804,Freitas:2016:ISS:3001610,10.1109/TASLP.2017.2738559,Chang2020-ks,Ghaffarzadegan2016-jf} has previously been conducted; however, to the best of the author’s knowledge, the use of mixing normal voices and whispering, has not considered.

The Alexa smart speaker supports the {\it whisper mode}~\cite{ai-alexa}. When this mode is set, if you talk to Alexa in a whisper, it will respond in a whisper~\cite{Cotescu2019-fr}.

\section{Interacting with DualVoice}

Figure~\ref{fig:sample} shows examples of voice input process using DualVoice.

In regular text creation by speech input, the voice input is converted into text by speech recognition when one speaks. However, when a misrecognition occurs, it is necessary to correct the input using a keyboard or other manual means. In addition, symbols such as periods, commas, quote marks, etc., need to be manually input. For example, saying ``period'' could mean you want to input the corresponding symbol, but it could also mean you want to input the actual word ``period' in letters. It is also difficult to use voice input for things that require command key input in normal text editing tasks, such as line breaks and paragraph changes.

In DualVoice, normal voice input is converted into text, the same as in conventional voice input text creation; whereas, symbols such as ``\whisp{COMMA},'' ``\whisp{PERIOD},'' and ``\whisp{QUOTE},'' in contrast, are input as symbols when spoken in a whisper. Whispering ``\whisp{NEW}\ \whisp{LINES}'' is also treated as a command and means making a new line. If there is a misrecognition, it is possible to delete the last word or change the word of interest by whispering ``\whisp{BACK}'' or ``\whisp{DELETE} \whisp{SENTENCE}.'' Presenting alternative recognition candidates is also possible by whispering ``\whisp{MENU}.'' Candidates on a menu are labeled as 1, 2, 3 and can be selected by whispering ``\whisp{ONE},'' ``\whisp{TWO},'' ``\whisp{THREE},'' etc. (Figure~\ref{fig:sample} (a, b)).

Combining normal voice input with whispered voice commands is also feasible.  For example, saying ``w a v 2 v e c'' and whispering ``\whisp{SPELL}'' would yield a ``wav2vec'' word, which is difficult to input with normal ASR (Figure~\ref{fig:sample} (c)).  Similarly, {\it Emoji} input is also possible by saying ``smile'' followed by whispering ``\whisp{EMOTION}'' (Figure~\ref{fig:sample} (d)).

\add{ 
The aforementioned commands show one possibility of DualVoice; however, as many function or control keys can be combined to represent many possible interactions, there are many possibilities of command configurations that utilize whisper discrimination and are worth exploring.
}

\section{Recognition Architecture}

This section describes the system architecture and neural network configurations that enable whisper voice recognition and classification. The overall architecture is shown in Figure~\ref{fig:wav2vec2asr}.

\subsection{Whisper Voice Recognition}

\begin{figure}
\centering
\includegraphics[width=0.47\textwidth]{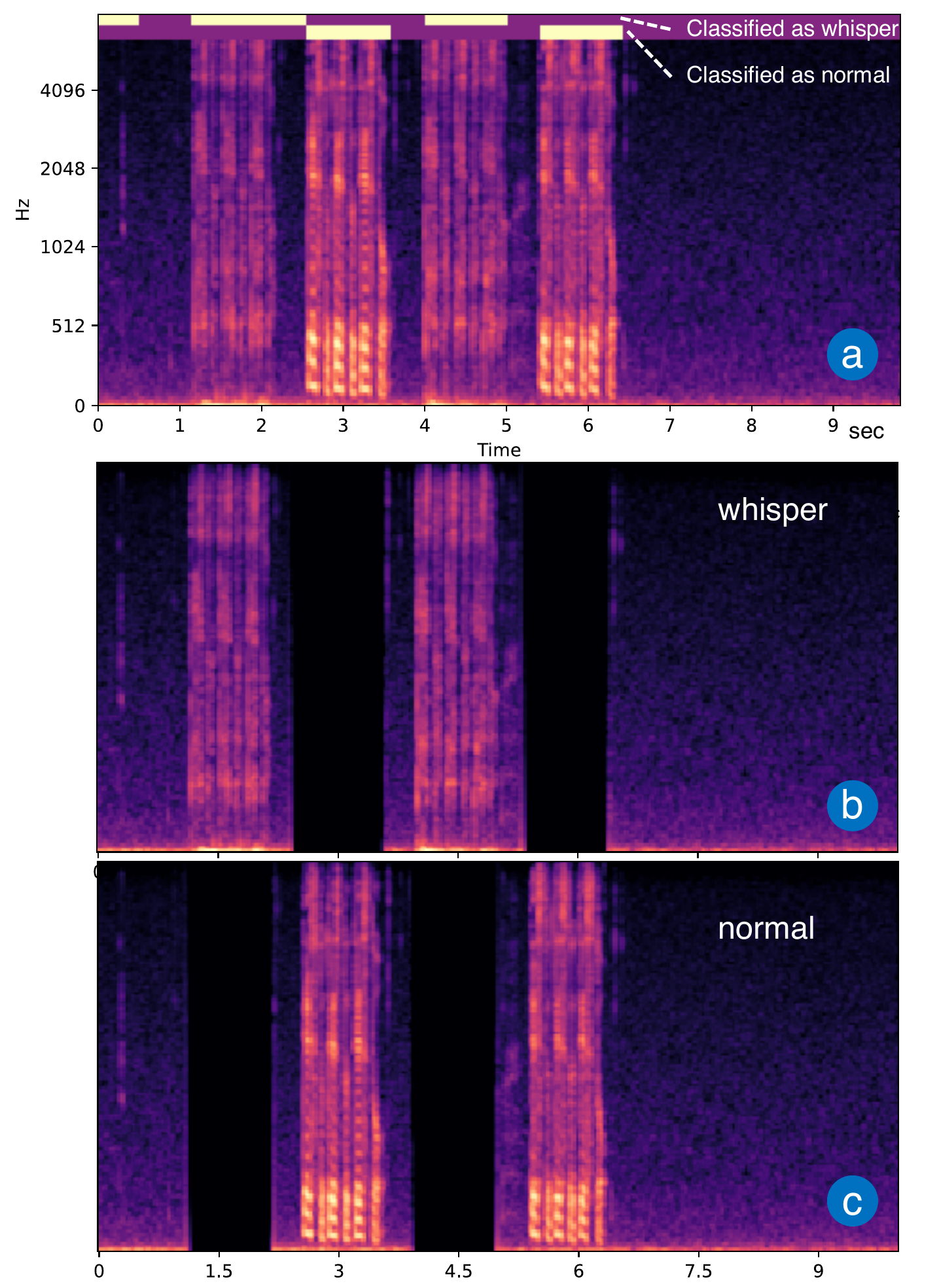}
\caption{Whisper classification examples: 
(a) the classification result, (b, c) the filtering results to be sent to the corresponding whisper/normal speech recognizers.}
\label{fig:whisperClassify}
\Description[Whisper classification examples]{Whisper classification examples}
\end{figure}

\begin{figure}
\centering
\includegraphics[width=0.47\textwidth]{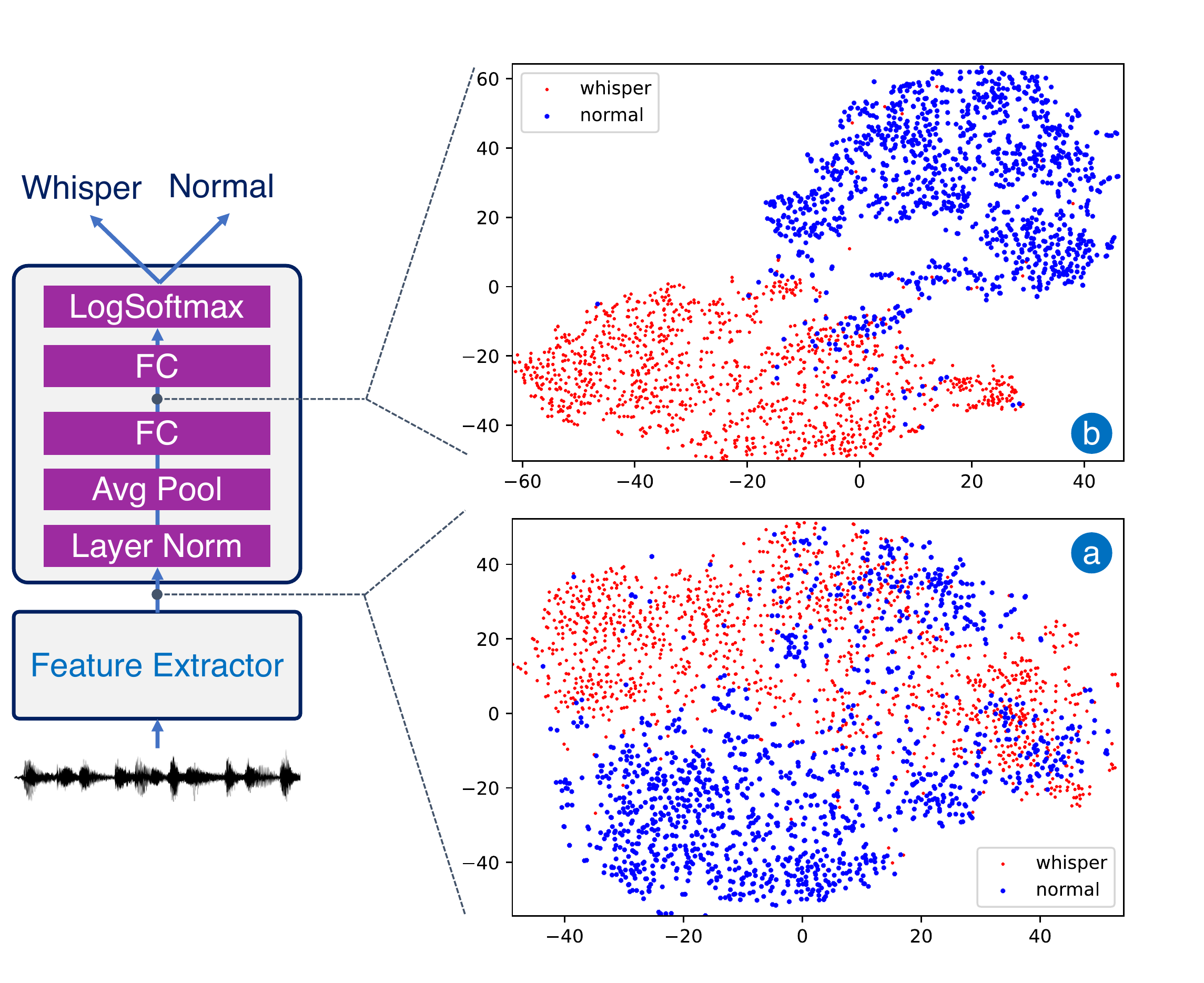}
\caption{t-SNE 2D visualizations of feature vectors: (a) output of the feature extractor, (b) vectors to be classified at the last FC (fully-connected) layer of the whisper classifier. It can be seen that whisper and normal voices are well separated. }
\Description[visualization of feature vectors]{t-SNE 2D visualizations of feature vectors}
\label{fig:tsne}
\end{figure}

For whispered voice recognition, this study utilized wav2vec 2.0~\cite{wav2vec2,Yi2020-re} and HuBERT (Hidden-Unit BERT)~\cite{10.1109/TASLP.2021.3122291}; both are self-supervised neural networks designed for speech processing systems.

Both networks assume a combination of pre-training with self-supervised representation learning on unlabeled speech data and fine-tuning on labeled speech data. These systems are primarily intended for speech recognition applications but have also been applied to speaker, language, and emotion recognition~\cite{Pepino2021-hv,Yi2020-re}.

The overall structure is shown in Figure~\ref{fig:wav2vec2asr}. It can be divided into the feature extractor and transformer layers~\cite{Vaswani2017-bm}. 

The pre-training method is similar to that for BERT (Bidirectional Encoder Representations  from Transformers)~\cite{Devlin2018-cg}'s masked language model in natural language processing. It is designed to mask a part of the input and estimate the expression features corresponding to it from the rest of the inputs. With this pre-training, the model can learn the acoustic properties of the input data and the characteristics of the speech.

In contrast to the pre-training, fine-tuning, which requires text labels for the audio data, requires only a small amount of data. As shown in Figure~\ref{fig:wav2vec2asr}, a projection layer and a connectionist temporal classification (CTC) layer~\cite{10.1145/1143844.1143891} were added to the output of wav2vec~2.0 or HuBERT to enable text transcriptions to be generated from audio waveforms.

As reported in~\cite{wav2vec2,Yi2020-re}, wav2vec~2.0 can achieve a speech recognition accuracy comparable to conventional state-of-the-art ASRs with fine-tuning only on a short amount of labeled speech dataset. Therefore, it is anticipated that this architecture is suitable for recognizing whisper voices under limited whisper speech corpora.

There are many corpora of normal speech; however, the corpora of whisper speech is limited, such as wTIMIT~\cite{wTIMIT}. Therefore, this study adopted a fine-tuning policy with whisper speech for neural networks pre-trained with normal speech corpora.

The feature extractor converts raw waveforms into latent feature vectors. Similar to the original wav2vec~2.0 and HuBERT, it consists of seven blocks and the temporal convolutions. Each block has 512 channels with strides (5,2,2,2,2,2,2) and kernel width (10,3,3,3,3,2,2) (Figure~\ref{fig:wav2vec2asr} (left)).
It is designed to output 512 latent dimension vectors every 20 ms.

\subsection{Whisper Voice Classification}

The whisper voice classification part distinguishes whispers from normal voice input by a fixed-length (e.g., 100 ms) audio signal, whereas the feature extractor based on convolutional neural networks obtained from wav2vec~2.0 converts the acoustic signal into 512-dimensional features every 20 ms  (Figure~\ref{fig:wav2vec2asr} (right)). As the whisper classification part could partially share the neural network with the whisper recognition part, the overall network size is reduced.

Using these features, the layer normalization, average pooling layers, followed by two FC (fully connected) layers, are applied to obtain whisper and normal voice classification.

The results of the classification sample are shown in Figure~\ref{fig:whisperClassify}. First, an audio stream with removed whispered voices and an audio stream with removed normal voices are generated from this information. Then, each speech stream is sent to its respective speech recognizer.

Figure~\ref{fig:tsne} shows the 2D plotting based on t-SNE~\cite{JMLR:v9:vandermaaten08a} of feature vectors to be classified as the input to the last FC (fully-connected) layer; the figure indicates that normal speech and whisper voice are well discriminated.

\add{Other whisper voice classification techniques based on acoustic analysis such as LPC residuals can also be used~\cite{ITO2005139,zhang07d_interspeech,10.1145/3271553.3271611}; one advantage of our approach is that the feature extractor part (CNN layers) can be shared with whisper voice recognition (Figure~\ref{fig:wav2vec2asr}). Only a relatively small network layer (with 260K parameters) need to be added for whispered speech classification, with this feature contributing to the simplicity of the system.}

\subsection{Training Dataset}

\begin{figure*}
\centering
\includegraphics[width=0.8\textwidth]{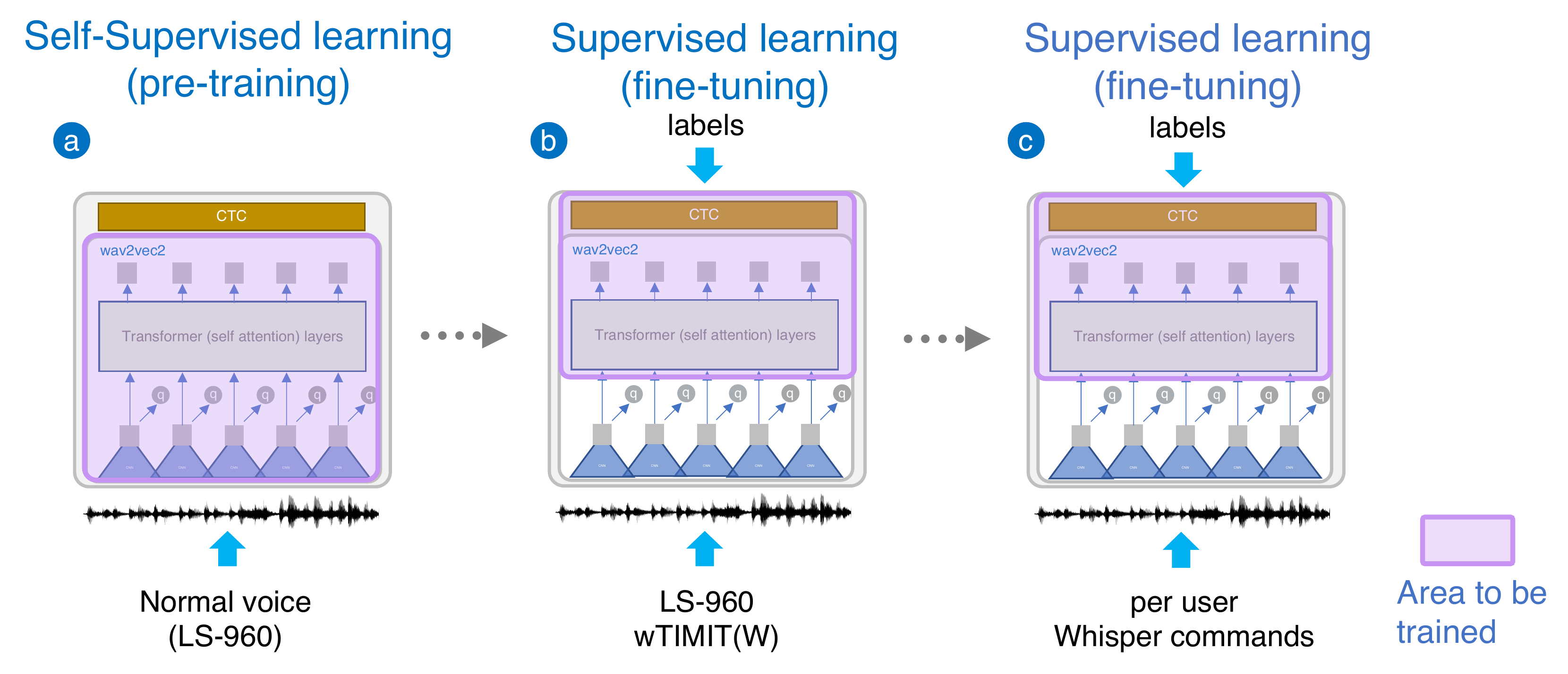}
\caption{Whisper speech recognition training process.}
\label{fig:wav2vec2_train}
\Description[Whisper speech recognition training]{Whisper speech recognition training process.}
\end{figure*}

\begin{table}
\centering
{\it
\begin{tabular}{ll}\toprule
one, two, three, four, five & six, seven, eight, nine, zero\\
A, B, C, D, E, F, G & H, I, J, K, L, M, N\\
O, P, Q, R, S, T, U &V, W, X, Y, Z\\
newline, back, delete &delete sentence\\
space, equal, at, number & dollar, ampersand, asterisk\\
left parenthesis, right parenthesis & left bracket, right bracket\\
underline, hyphen, plus, minus & percent, atmark, sharp\\
spell, paragraph & period, comma, dot\\
menu, open, close, yes, no & line, new, repeat, candidates\\
next, page, word & delete line, delete word\\
question mark, exclamation mark & quote, double quote\\\bottomrule
\end{tabular}
}
\caption{Whisper phrases used for per-user training.}
\label{tab:peruser}
\Description[Whisper phrases]{Whisper phrases used for per-user training}
\end{table}

\begin{table*} 
\begin{tabular}{l|ll|rr}\toprule
model & train \add{(fine-tuning)} & test & WER (\%) & CER (\%) \\\midrule
\add{Google Text-to-speech~\cite{googlespeech}} & & \add{wTIMIT(N)} & \add{11.55} & \add{4.66} \\
                      & & \add{wTIMIT(W)} & \add{44.70} & \add{28.38} \\\midrule
wav2vec 2.0 large~\cite{wav2vec2} & LS-960 & wTIMIT(N) & 15.85 & 4.42 \\
& LS-960 & wTIMIT(W) & 39.00 & 16.46 \\
& LS-960 + wTIMIT(W) & wTIMIT(W) & 0.84 & 0.45 \\
& LS-960 & per-user & 83.95 & 38.42 \\
& LS-960 + wTIMIT(W) & per-user & 53.33 & 34.64 \\
& LS-960 + per-user & per-user & {\bf 2.45} & {\bf 0.56}\\
& LS-960 + wTIMIT(W) + per-user & per-user & {\bf 0.38} & {\bf 0.18} \\\midrule
HuBERT large~\cite{10.1109/TASLP.2021.3122291} & LS-960 & wTIMIT(N) & 11.03 & 2.76 \\ 
& LS-960 & wTIMIT(W) & 17.12 & 5.74 \\
& LS-960 + wTIMIT(W) & wTIMIT(W) & 0.78 & 0.43 \\
& LS-960 & per-user & 61.24 & 25.34 \\
& LS-960 + wTIMIT(W) & per-user & 83.25 & 37.32 \\
& LS-960 + per-user & per-user & {\bf 0.34} & {\bf 0.08} \\
& LS-960 + wTIMIT(W) + per-user & per-user & {\bf 0.32} & {\bf 0.06} \\\bottomrule
\end{tabular}
\caption{Whisper voice recognition accuracy for \add{Google Cloud Speech-to-Text}, wav2vec~2.0 large, and HuBERT large: The latter two models are pre-trained on unlabeled LibriSpeech for 960 hours with (LS-960) audio data and fine-tuned on the same LS-960 with text labels. The ``train'' column indicates dataset used for fine-tuning (LS-960: LibriSpeech 960 hours, wTIMIT(N): wTIMIT normal voice, wTIMIT(W): wTIMIT whisper voice, and per-user: per-user whisper voice). For wav2vec~2.0 and HuBERT models, scores fine-tuned with wTIMIT(W) and per-user were the best; however, that fined-tuned with only per-user data (and not with wTIMIT(W)) also showed comparable results.}
\label{tab:wer}
\Description[Whisper voice recognition accuracy]{Whisper voice recognition accuracy}
\end{table*}

%Following the above frameworks, 
The neural networks were pre-trained and fined-tuned with (normal) voice data (960 hours from the LibriSpeech dataset~\cite{librispeech}). Notably, the pre-training did not require the corresponding text, only the voice data (Figure~\ref{fig:wav2vec2_train}). The following two whispered voice datasets were then used for further fine-tuning:

\subsubsection*{wTIMIT}
wTIMIT whisper voice corpus ({\it whisper TIMIT})~\cite{wTIMIT}. Each speaker both says and whispers 450 phonetically balanced sentences according to the prompts in TIMIT, with 29 speakers that are gender-balanced to some extent. The total number of utterances used is 11,324 (1,011 minutes); it also comes with a normal voice. These were used for whisper/normal classification training.
The data is provided in two parts, train and test, with the division kept as is.

\subsubsection*{Per-user dataset}
Voice data dubbed with whisper voice by each user according to the prompts of the selected voice phrases (Table~\ref{tab:peruser}). These phrases are mainly assumed to be the commands used during text input. Each user repeats each phrase five times, for a total of 110 phrases \add{(average 2.7 seconds). The total data acquisition time for each user is approximately 15 minutes. } These phrases are then randomly concatenated to be used as a dataset. After this concatenation, the total number of utterances is 936 (approximately 82 minutes).

Fine-tuning is applied in two stages. In the first stage, the networks are trained on the wTIMIT (whisper voice), and in the second stage, the whispering voice command set blown by the users (per-user dataset) is used. 

The wTIMIT normal and whisper voice data is also used as the training dataset for the training of whisper/normal classification. The length of the voice supplied to the classifier is set to 1,600 samples (100 ms in size with $16\  K / sec$ audio sampling), which is consistent with the length of the speech chunks used in the speech recognition cloud service in the later stage.

\section{Results}

\subsection{Whisper voice recognition: }

\begin{table} 
\begin{tabular}{l|ll|r}\toprule
model & train & test & acc. \\\midrule
feature extractor (Fig.~\ref{fig:wav2vec2asr}) + & wTIMIT & wTIMIT & {\bf 0.967} \\
\,LayerNorm + Pooling + FC + FC                  & wTIMIT & per-user & 0.905  \\
\midrule                 
MFCC +  & wTIMIT & wTIMIT & 0.949 \\
\,LayerNorm + Pooling + FC + FC     &  wTIMIT & per-user & 0.857 \\ 
\bottomrule
\end{tabular}
\caption{Whisper--normal classification accuracy results.}
\label{tab:whisperclass}
\Description[Whisper-Normal classification]{Whisper-Normal classification accuracy results.}
\end{table}

The results of the training are shown in Table~\ref{tab:wer}. The recognition scores are shown as word error rate (WER) and character error rate (CER). The training times (using a single NVIDIA A6000 GPU) for fine-tuning were 4.5 hours for wTIMIT(W) and 20 minutes for the per-user dataset.

The recognition accuracy was determined to be poor when trained (fine-tuned) with LS-960 and wTIMIT(W) and tested with the per-user dataset, indicating that this neural network is not sufficient for an unspecified speaker whisper recognizer. In contrast, the system pre-trained on LS-960 and fine-tuned with LS-960, wTMIT(W), and per-user datasets showed good recognition accuracy (e.g., WER 0.38\%).

These results indicate that (1) the model pre-trained on normal speech can deliver good whisper recognition accuracy when it is fine-tuned with whisper voice, and (2) at this stage, only a user-dependent whisper recognition is achievable.

\subsection{Whisper--Normal Classification: }

For the whisper--normal classifier training, this study utilized the normal and whisper voices of wTIMIT. The 16-kHz sampled voice dataset was divided into an audio segment containing 1,600 samples (100 ms). Each segment was labeled either normal or whisper. The system was then trained to discriminate between these two.

Of the 100 ms speech segments, those with an average power of less than -20 db were excluded, and training was performed on the rest. The results are shown in Table~\ref{tab:whisperclass}). As can be seen in the table, classification of normal and whispered speech was possible with a 96.7 \% accuracy. 

A comparison of models that use feature extractor to those that use MFCC (Mel-Frequency Cepstrum Coefficient) as a feature was also conducted. As shown in the table, the model using feature extractor outperformed the model using MFCC slightly. In particular, the feature extractor based model performed better in discrimination tested by the voice of a user different from wTIMIT.

\subsection{Normal Voice Recognition}

For normal voice recognition, although the afore-described neural network could be trained, this study opted to select an existing cloud-based speech recognition system (Google Cloud Speech-to-Text~\cite{googlespeech}) because of the noted better recognition accuracy and compatibility with existing voice-based text input applications.

\begin{figure}
\centering
\includegraphics[width=0.47\textwidth]{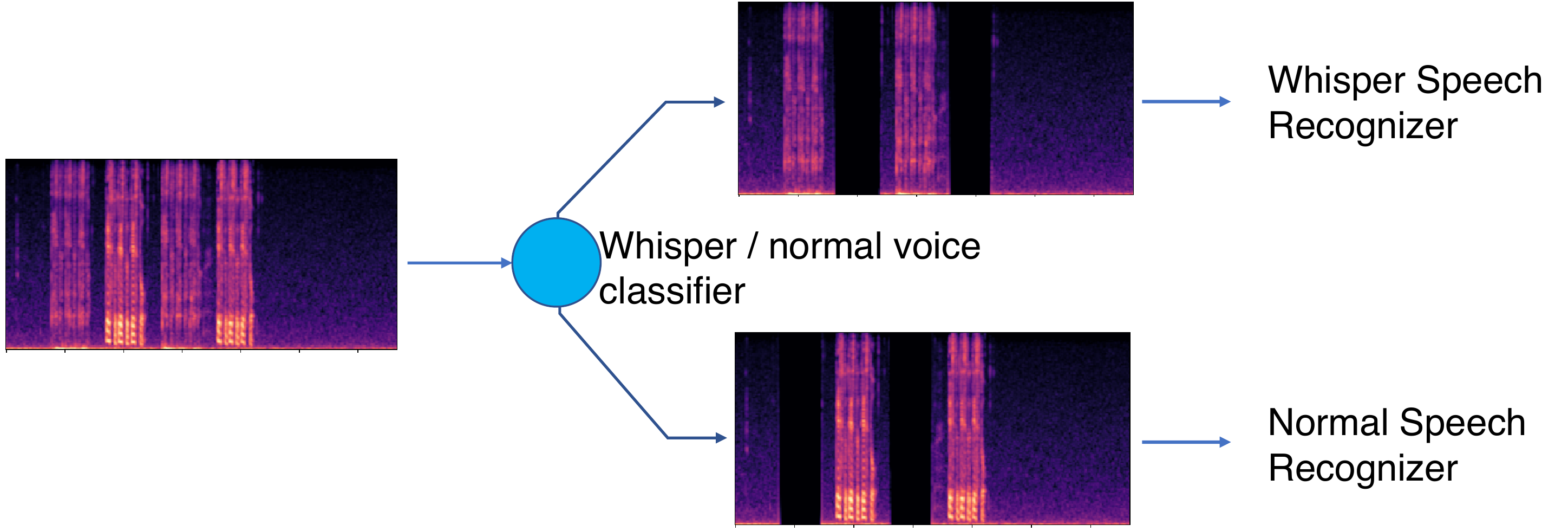}
\caption{Whisper classification and filtering: As a result of whisper--normal classification, the speech stream with the whisper part removed is sent to the normal speech recognizer, and the speech stream with the normal part removed is sent to the whisper speech recognizer.}
\label{fig:filter}
\Description[Whisper classification and filtering]{Whisper classification and filtering}
\end{figure}

\begin{figure*}
\centering
\includegraphics[width=0.9\textwidth]{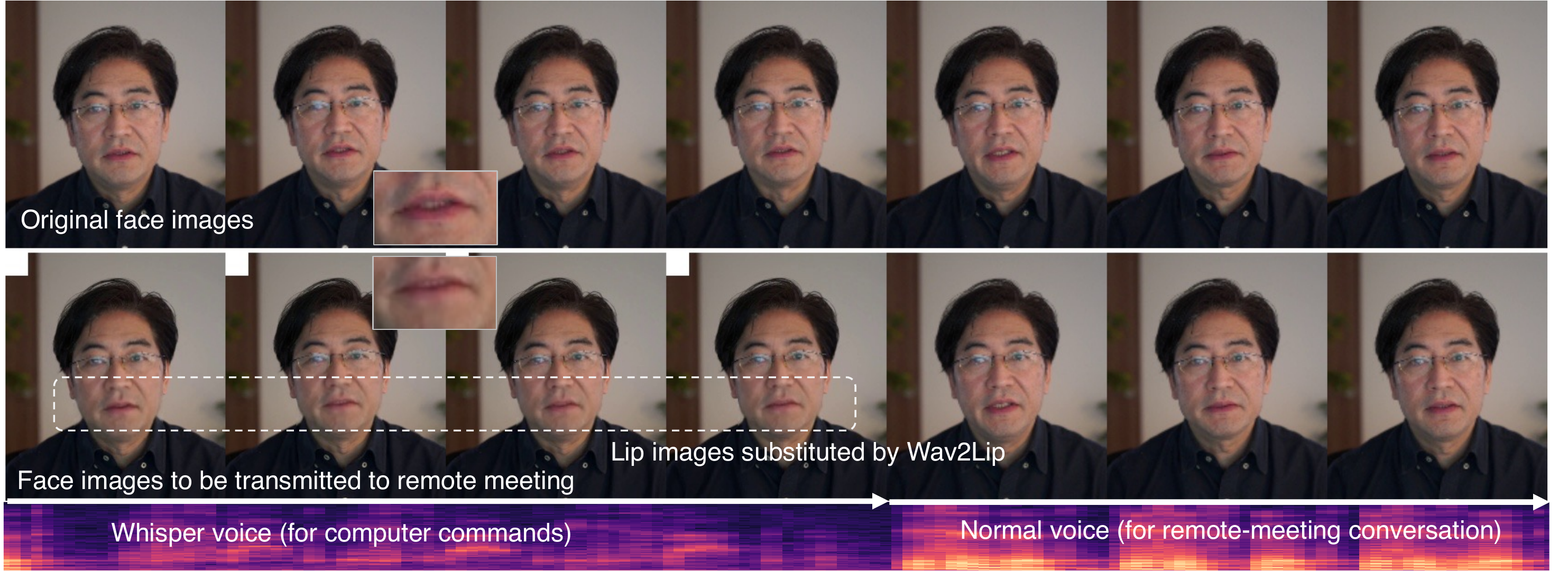}
\caption{Giving commands to the computer with whisper voice while using normal voice for remote meeting conversation. While speaking in a whisper voice, lip images are substituted using Wav2Lip~\cite{wav2lip}, so that the mouth appears to be closed during whisper speech. \add{(note: this is a partially envisioned example; Lip image substitution is realized offline, whereas whisper identification is done in real time.)}}
\label{fig:mouth}
\Description[Giving commands as whisper voice during a meeting]{Giving commands to the computer with whisper  voice while using normal voice for remote meeting}
\end{figure*}

\section{Implementation Details}

The neural networks for whisper recognition and whisper discrimination are based on huggingface~\cite{Wolf2019-vq} wav2vec~2.0 and HuBERT classes, and other networks developed with PyTorch~\cite{paszke2017automatic}. The GUI system for the DualVoice text input example is built using the Tkinter Python GUI platform~\cite{lundh1999introduction}.

The GUI system manages the thread that receives the microphone input and splits it into packets containing 1,600 samples ($100 ms$) of sound. These packets are sent via TCP/IP to the whisper discriminator, which discriminates each packet as containing whispers, normal voice, or silence. 
Then, depending on the result of the discrimination, the speech packet is then sent to either the whisper recognizer or the normal speech recognizer. 

In the case of speech sent to whisper speech recognition, packets that are judged not to be whisper speech will be replaced with no-speech packets; in the case of normal speech recognition, packets that are classified to be whisper speech will be replaced with no-speech packets (Figure~\ref{fig:filter}).

\begin{figure}
\centering{\includegraphics[width=0.45\textwidth]{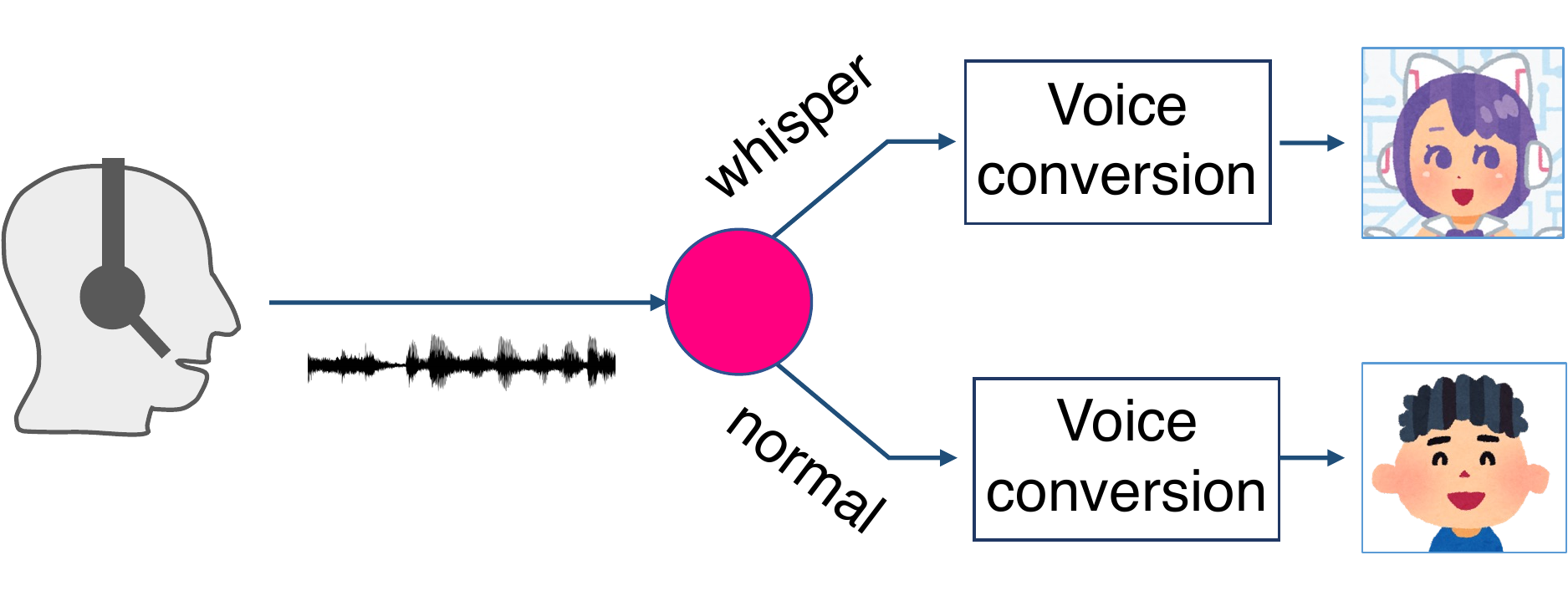}}
\caption{Controlling multiple avatars with normal and whisper voice.}
\label{fig:avatar}
\Description[Controlling multiple avatars]{Controlling multiple avatars with normal and whisper voice}
\end{figure}

\begin{figure}
\centering
\includegraphics[width=0.4\textwidth]{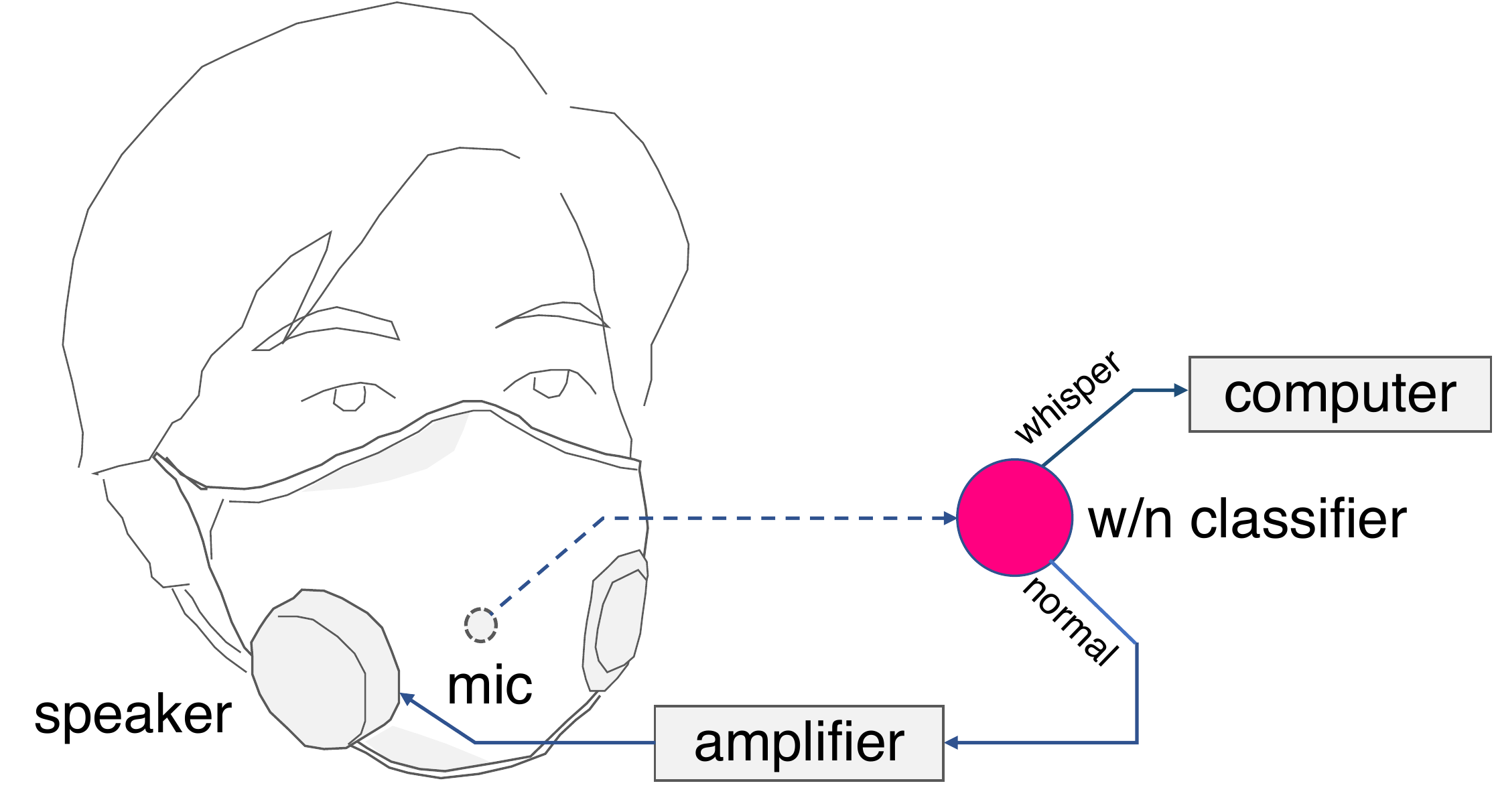}
\caption{Masked system with whisper discrimination function: Normal voice is output externally via speaker, and whisper voice is treated as a command to the computer.}
\label{fig:mask}
\Description[Masked system]{Masked system with whisper discrimination function}
\end{figure}

\section{End-User Experience}

Five participants were asked to use the DualVoice system for speech input. The whisper voice recognition neural network was first fine-tuned for each user and then the participants were asked to enter text sentences verbally. 

The use of whisper speech as a command was understood and achievable by all users. \add{All participants were observed to be able to easily distinguish between normal and whisper utterances, although a very short momentary pause was usually required to switch between the normal and whispered voice. }

Participant feedback indicated that it was slightly inconvenient to learn which commands could be executed with a whisper voice and suggested that the design guidelines might vary depending on whether the goal is to operate the system entirely by voice alone or to use it in conjunction with a mouse or keyboard. There was also a suggestion that the system could be used as an input method for people with paralyzed fingers (without the need for keyboard or mouse manipulations).

\section{Discussions}

This section briefly describes the potential and current issues of this method.

\subsection{Other Potential Applications}

The whisper voice discrimination enables multiple modes of speech interaction, and the possibilities are not limited to text input. Some possible applications are as follows.

\subsection*{Using whispering commands at a teleconference}

DualVoice can be used during teleconferences.

Assume the use of voice commands during a teleconference. The command is given in a whisper to prevent the commands from becoming part of the conference speech. The whisper discriminator discriminates between whisper and normal voices, and the whisper voice part is removed from the conference participant's audio stream. The whisper part is sent to the whisper recognizer.

As it is unnatural if the mouth is moving while uttering whisper speech; hence, the lip part should be replaced to appear as if the participant is not speaking (Figure~\ref{fig:mouth}). For this purpose, Wav2Lip can generate lip images from vocalizations using deep learning~\cite{wav2lip}. When the user speaks in a whisper voice, the face image is replaced by an image of the user when not speaking (i.e., a closed mouth).

Using this mechanism, both the face image and the voice stream can be adjusted so that the part where the command is uttered in a whisper voice would be invisible and inaudible to other participants.

Figure~\ref{fig:mouth} shows the original face image video and the face image when the part of whisper utterance is replaced with no utterance.

\subsection*{Control of multiple Avatars}

Switching the speech of multiple avatars is possible in virtual space (Figure~\ref{fig:avatar}). For example, the first avatar can be made to speak in a normal voice, and the second avatar could be made to talk in a whisper voice. Whisper voice utterances can then be expected to be converted to other voices using voice conversion technology~\cite{arxiv.1808.10687,8835014,arxiv.2004.09347}.

\subsection*{Combination with Silent Speech}

The objectives of silent speech are to ensure that the speech for speech command does not become noise to the surroundings; moreover, it preserves privacy by not disclosing confidential information~\cite{Denby:2010:SSI:1746726.1746804,Freitas:2016:ISS:3001610}. The sound pressure level of a typical conversation is approximately 60 dB; whereas, the sound pressure level of a whisper is in the range 30--40 dB. Thus, using a whisper voice as the speech command, the objectives of silent speech can be achieved.

Incorporation of whisper functionality into a mask-type wearable interface is also possible (Figure~\ref{fig:mask}). Philips and Dyson, for example, are in the process of developing masks that allow powered ventilation for breathing to protect against air pollution and infectious diseases~\cite{pmask,dysonzone}. A microphone can be placed inside such masks 
for picking up whisper voice, making it possible to achieve an effect almost equivalent to silent speech. By introducing a whisper discrimination function into such always-worn masks, an always-available voice interface without interfering normal conversation would be constructed.

Furthermore, there is potential to use three modalities: silent, normal, and whispered speech. For example, if lip reading and whisper speech can be recognized, three types of speech modality can be obtained together with normal speech.

\subsection{User-Independent Whisper Recognition} 

In the implementation presented in this study, an existing speech recognition system (Google Cloud Speech-to-Text~\cite{googlespeech}) was used for normal voice recognition and our custom neural network for whisper voice recognition. The former can be trained on a large corpus, is speaker-independent, and can be used without special training. In contrast, the whisper speech recognition part can be trained on a smaller corpus. It does require a training phase in which individual users are asked to read out example sentences in a whisper for speaker adaptation.

As whisper-based interactions become more widespread, it is anticipated that obtaining a whisper-speech dataset from the usage history will become possible. Hence, one will be able to use whisper speech recognition in a speaker-independent manner.

\section{Conclusion}

This study proposed DualVoice, a speech input method for inputting non-text commands in a whispered voice and inputting text in a normal voice. Normal voice input is used for text input, whereas, various commands can be entered by whispering them. The proposed method does not require any specialized hardware other than a regular microphone and can be used in a wide range of situations where speech recognition is already available.

Furthermore, this study designed two neural networks, one for distinguishing whisper speech from normal speech and the other for recognizing whisper speech and implemented a prototype speech-based text input system using these neural networks and evaluated its usability.

\begin{acks}
The authors wish to thank the anonymous reviewers for their valuable comments on earlier version of this paper.
This work was supported by JST Moonshot R\&D Grant Number JPMJMS2012, JST CREST Grant Number JPMJCR17A3, and The University of Tokyo Human Augmentation Research Initiative.
\end{acks}

\bibliographystyle{ACM-Reference-Format}
\bibliography{reference}

\end{document}